\documentclass[11pt,a4 paper]{article}

\usepackage{amssymb}
\usepackage{amsmath}
\usepackage[dvips]{graphicx}
\begin{document}
\thispagestyle{empty}
\begin{center}
\noindent {\textbf{\large Some Exact Solutions of Magnetized Viscous Model in String Cosmology}}\\

\vspace{0.5cm}
\noindent \textbf{C.P. Singh $^a$,\; Vijay  Singh$^b$ } \\

\vspace{0.5cm}

\noindent{Department of Applied Mathematics,\\
 Delhi Technological University (Formerly Delhi College of
 Engineering)\\
 Bawana Road, Delhi-110 042, India.}\\
$^a$ E-mail: \texttt{cpsphd@rediffmail.com}\\
$^b$ E-mail: \texttt{gtrcosmo@gmail.com}
\end{center}

\vspace{1.5cm}

\noindent\textbf {Abstract.} In this paper, we study anisotropic Bianchi-V universe with magnetic field and bulk viscous fluid in string cosmology. Exact solutions of the field equations are obtained by using the equation of state for a cloud of strings, and  a relationship between bulk viscous coefficient and scalar expansion. The bulk viscous coefficient is assumed to be inversely proportional to the expansion scalar. It is interesting to examine the effects of magnetized bulk viscous string model in early and late stages of the evolution of the universe. This paper presents the different string models like geometrical (Nambu string), Takabayashi (p-string) and Reddy string models by taking certain physical conditions. We discuss the nature of the classical potential for viscous fluid with and without magnetic field. We find that the introduction of bulk viscosity with and without magnetic field results in rapid change in density parameters and in the classical potential. The presence of bulk viscosity prevents the universe to be empty in its future evolution. The other physical and geometrical aspects of each string model are discussed in detail.\\

\noindent PACS number(s): 98.80-k, 98.80-cq, 04.20-q.\\
\pagebreak

\pagestyle{myheadings}
\begin{center}
\noindent \textbf{1. Introduction}\\
\end{center}

 \indent Recently, the string cosmology has received considerable attention into the framework of general relativity mainly because of its possible role in the early universe. The concept of string cosmology was developed to describe the events at the very early stages of the evolution of the universe. The presence of the strings during the early universe can be explained using grand unified theories [1-3]. In the early stages of the evolution of the universe it is expected that topological defects could have formed naturally during the phase transitions followed by spontaneous broken symmetries. Cosmic strings are linear topological defects, have very interesting properties and might play an important role in the structure formation. These cosmic strings have stress energy and coupled to the gravitational field. The gravitational effects of string in general relativity have been studied by Letelier [4] and Stachel [5]. Letelier [6] studied relativistic cosmological solutions of cloud formed by massive strings in Bianchi type-I and Kantowski-Sachs space-times. In these models, each massive string is formed by a geometric string with particles attached along its extension. In principle, the string can be eliminated and be ended up with the cloud of particles. This is a desirable property of a model of a string cloud to be used in cosmology. Matraverse [7] presented a class of exact solutions of Einstein field equations with a two parameter family of classical strings as the source of the gravitational field. Exact solutions of string cosmology in Bianchi type II, $VI_0$, VIII and IX space-times have been studied by Krori et al. [8]. Yavuz and Tarhan [9], Bali and Dave [10, 11], Bali and Upadhaya [12], Bali and Singh [13], Bali and Pradhan [14], Pradhan and Chouhan [15], Mahanto et al. [16] have investigated Bianchi type string cosmological models in general relativity. \\
 \indent The study of magnetic field provides an effective way to understand the initial phases of the cosmic evolution. Primordial magnetic field of cosmological origin have been discussed by Asseo and Sol [17], and Madsen [18]. Wolfe et al. [19], Kulsrud et al. [20], Barrow [21] have studied the cosmological models with magnetic field and have pointed out its importance in the early evolution of the universe. Matravers and Tsagas [22] have found that the interaction of the cosmological magnetic field with the space-time geometry could affect the expansion of the universe. Banerjee et al. [23], Chakraborty [24], Tikekar and Patel [25, 26] have studied Bianchi type I and III string cosmological models with and without source-free magnetic field. ShriRam and Singh [27] have obtained some new exact solutions of string cosmology with and without  a source-free magnetic field in Bianchi type I space-time in different basic forms. Patel and Maharaj [28] and Singh and Singh [29] have studied string cosmology with magnetic field in anisotropic models. Singh and ShriRam [30] have presented a technique to generate new exact Bianchi type III string cosmological solutions with magnetic field. Kilin\c{c} and Yavuz [31], Pradhan et al. [32], Pradhan [33], Bali and Jain [34], Saha and Visinescu [35], Pradhan et al. [36, 37], Saha et al. [38], Pradhan et al. [39], Amirhashchi et al. [40] and Rikhvitsky et al. [41] have investigated string cosmological models in the presence and absence of magnetic field. \\
 \indent The observations indicate that the media is not a perfect fluid and the viscosity is concerned in the evolution of the universe [42]. Bulk viscosity has essential importance in early universe. In order to study the evolution of the universe, many authors [43-51] have investigated cosmological models with a fluid containing viscosity. Yadav [52], Mohanty and Gauranga [53], Tripathy and Behara [54] are some of the authors who have studied string cosmological models with viscous fluid. \\
 \indent Bianchi type-V space-times are interesting to study the evolution of universe because of their richer structure both physically and geometrically than standard Freidmann-Roberson-Walker (FRW) models. These models represent the open FRW cosmological model. Coley [55], Singh et al. [56-58], Singh and Beesham [59], Singh [60] have studied Bianchi V models in general relativity in many physical contexts. Chakraborty and Chakraborty [61], and Bali [62] have studied string cosmology with magnetic field in Bianchi V space-time model for different directions of the magnetic field and string. Some authors [63, 64] have investigated anisotropic models with magnetic field and bulk viscosity in string cosmology. Recently, Sharif and Waheed [65], and Singh [66] have studied Bianchi I magnetized viscous fluid model in string cosmology and have discussed the effect of viscous fluid and magnetic field on the classical potential. Therefore, the study of magnetized cosmic strings with bulk viscosity in richer structure of an anisotropic model like Bianchi V leads to a better understanding of the dynamics of the universe during early and late stages of the evolution. \\
 \indent The aim of this paper is to find some exact solutions in Bianchi V model with magnetized viscous fluid in string cosmology and discuss their effects in early and late time evolution of the universe. The equation of state for a cloud of strings, and the inverse relation between expansion scalar and bulk viscous coefficient are used to solve the field equations exactly. The paper is organized as follows.\\
 \indent In section 2 the field equations for Bianchi V model with bulk viscosity and magnetic field in string cosmology are presented. Section 3 deals with the solution of field equations where a general quadrature form of volume scale factor of the model is given. Three different exact string models are presented in subsections 4.1, 4.2 and 4.3 of section 4. We discuss the various physical parameters in each case.  Finally, we present a summary of the results in section 5.\\

\begin{center}
\noindent\textbf{2. Model and Basic Equations}\\
\end{center}
\noindent We consider the homogenous and anisotropic Bianchi -V metric in the form
\begin{equation}
ds^{2} =-dt^{2}+A^2dx^2+e^{2x}\left(B^2dy^{2}+C^2dz^2\right),
\end{equation}
where $A$, $B$, and $C$ are scale factors in anisotropic background and are the functions of cosmic time $t$ only.\\
\indent The energy-momentum tensor for bulk viscous string dust with magnetic field is given by [6, 67]
 \begin{equation}
  T^j_{i}=\rho u_iu^{\;j}-\lambda x_ix^j-\xi u^l;_l\left(g^j_i+u_iu^j\right)+E_i^{\;j},
\end{equation}
\noindent where $\rho$ is the proper energy density for a cloud of strings with particles attached to them and $\lambda$ is the string tensor density and  is related by $\rho=\rho_p+\lambda$, where $\rho_p$ is the particle energy density. The unit time-like vector $u^i$ describes the four velocity of the particle and unit space-like vector $x^i$ denotes the direction of the string which can be taken along any of the three directions $x$, $y$ and $ z$ axes. Thus, without loss of generality let us choose $x-$ direction as the direction of the string along which the magnetic field is assumed to be present, i.e.,
\begin{equation}
  x^i=(A^{-1},0,0,0)
\end{equation}
In a co-moving coordinate system, we have
\begin{equation}
  u^i=(0,0,0,1)
\end{equation}
Therefore, we have
\begin{equation}
u_iu^i=-x_ix^i=-1, \;\; u_ix^i=0 .
\end{equation}
\indent In Eq. (2), $\xi$ is the coefficient of bulk viscosity, $u^l_{;l}=\theta$ is the expansion scalar and $E_i^{\;j}$ is the electromagnetic field tensor which is given by (see, ref.[68])
\begin{equation}
E_i^{\;j}=\bar\mu \left[|h|^2\left(u_iu^j+\frac{1}{2}g_i^{\;j}\right)-h_ih^j\right],
\end{equation}
where $\bar\mu$ is the magnetic permeability, and $h_i$, the magnetic flux vector defined by
\begin{equation}
h_i=\frac{1}{\bar\mu}\bar F_{ij}u^j.
\end{equation}
\noindent The dual  electromagnetic field tensor $\bar F_{ij}$ is defined as
\begin{equation}
  \bar F_{ij}=\frac{\sqrt{-g}}{2}\epsilon_{ijkl}F^{kl}
\end{equation}
where $F^{kl}$ is the electromagnetic field tensor and $\epsilon_{ijkl}$ is the Levi-Civita tensor density.\\
\indent We assume that the magnetic field is generated in $yz$-plane as its source is the electric current that flows in $x-$ direction. Therefore, the magnetic flux vector has only one non-zero component $h_1$, i.e., $h_1\neq0$, $h_2=0=h_3=h_4$. Moreover, the assumption of infinitely electrical conductivity [69] along with finite current leads to  $F_{14}=0=F_{24}=F_{34}$.\\
\indent Using Maxwell's equations
\begin{equation}
F_{ij;k}+F_{jk;i}+F_{ki;j}=0, \;\;\text{and}\;\; F^{ij}_{;k}=0,
\end{equation}
we find
\begin{equation}
  F_{23}=I = constant.
\end{equation}
\noindent  Hence, the non-zero component of magnetic flux vector is
\begin{equation}
h_1=\frac{AI}{\bar\mu BC}.
\end{equation}
\noindent Since $|h|^2=h_lh^l=h_1h^1=g^{11}(h_1)^2$, therefore,
\begin{equation}
|h|^2=\frac{I^2}{\bar\mu^2B^2C^2}\;.
\end{equation}
\noindent Using Eqs. (11) and (12) into (6), the components of $E_{i}^{j}$ are given by
\begin{equation}
E^1_{\;1}=-\frac{I^2}{2\bar\mu^2 B^2C^2}=-E^2_{\;2}=-E^3_{\;3}=E^4_{\;4}.
\end{equation}
\noindent The Einstein's field equations (in gravitational units $c=8\pi G=1$) read as
\begin{equation}
  R_i^{\;j}-\frac{1}{2}g_i^{\;j}R=-T_i^{\;j},
\end{equation}
where $R_i^j$ is the Ricci tensor and $R=g^{ij}R_{ij}$ is the Ricci scalar.\\
\indent The field equations (14) with (1) and (2) subsequently lead to the following system of equations
\begin{equation}
  \frac{\dot A\dot B}{AB}+\frac{\dot B\dot C}{BC}+\frac{\dot C\dot A}{CA}-\frac{3}{A^{\;2}}=\rho+\frac{I^2}{2\bar\mu B^{\;2}C^{\;2}},
\end{equation}
\begin{equation}
\frac{\ddot A}{A}+\frac{\ddot C}{C}+\frac{\dot A\dot C}{AC}-\frac{1}{A^2}=-\frac{I^2}{2\bar\mu B^{\;2}C^{\;2}}+\xi\theta,
\end{equation}
\begin{equation}
\frac{\ddot A}{A}+\frac{\ddot B}{B}+\frac{\dot A\dot B}{AB}-\frac{1}{A^2}=-\frac{I^2}{2\bar\mu B^{\;2}C^{\;2}}+\xi\theta,
\end{equation}
\begin{equation}
\frac{\ddot B}{B}+\frac{\ddot C}{C}+\frac{\dot B\dot C}{BC}-\frac{1}{A^2}=\lambda+\frac{I^2}{2\bar\mu B^{\;2}C^{\;2}}+\xi\theta,
\end{equation}
\begin{equation}
2\frac{\dot A}{A}-\frac{\dot B}{B}-\frac{\dot C}{C}=0,
\end{equation}
\noindent where the over dots indicate ordinary differentiation with respect to $t$.\\
\indent Let us define the volume scale factor $\tau$ as
\begin{equation}
  \tau=ABC.
\end{equation}
\noindent Let us consider the various important  physical quantities such as expansion scalar $\theta$, anisotropy parameter $\Delta$ and shear scalar $\sigma^2$, which are defined as
\begin{equation}
\theta=u^l_{;l}=\frac{\dot A}{A}+\frac{\dot B}{B}+\frac{\dot C}{C}=\frac{\dot\tau}{\tau},
\end{equation}
\begin{equation}
\Delta=\frac{1}{3}\sum_{i=1}^3\left(\frac{H_i-H}{H}\right)^2,
\end{equation}
\begin{equation}
\sigma^2=\frac{1}{2}\;\sigma_{ij}\sigma^{ij}=\frac{1}{3}\Delta H^2.
\end{equation}
\noindent Here, $H=\theta/3$ is the mean Hubble parameter and $H_1=\frac{\dot{A}}{A}$, $H_2=\frac{\dot{B}}{B}$ and $H_3=\frac{\dot{C}}{C}$ are the directional Hubble parameters in the directions of $x$, $y$ and $z$ axes, respectively. \\
\indent The energy conservation equation $T^j_{i;j}=0$ takes the form
\begin{equation}
  \dot\rho+\frac{\dot\tau}{\tau}\rho-\frac{\dot A}{A}\lambda=\xi\frac{\dot\tau^2}{\tau^2}.
\end{equation}

\begin{center}
  \noindent\textbf{3. Solution of the field equations }
\end{center}
\indent To solve the field equations (15)-(19), we follow the method recently used by Saha and Visinescu [36]. From (16) and (17) we get
\begin{equation}
\frac{d}{dt}\left(\frac{\dot C}{C}-\frac{\dot B}{B}\right)+\left(\frac{\dot C}{C}-\frac{\dot B}{B}\right)\left(\frac{\dot A}{A}+\frac{\dot B}{B}+\frac{\dot C}{C}\right)=0.
\end{equation}
\noindent Using (20) into (25), we get
\begin{equation}
\frac{d}{dt}\left(\frac{\dot C}{C}-\frac{\dot B}{B}\right)+\left(\frac{\dot C}{C}-\frac{\dot B}{B}\right)\frac{\dot\tau}{\tau}=0,
\end{equation}
\noindent which on integration, gives
\begin{equation}
C=d_1B \exp\left({k_1\int\frac{dt}{\tau}}\right),
\end{equation}
\noindent where $d_1$ and $k_1$  are constants of integration.\\
\indent From (19), we get
\begin{equation}
A^2=d_2BC,
\end{equation}
\noindent  where $d_2$ is a constants of integration, which has been taken unity without loss of generality.\\
\indent From (15)-(18) and (28), we obtain
 \begin{equation}
\frac{\ddot\tau}{\tau}=\frac{1}{2}\left[(3\rho+\lambda)+\frac{A^2I^2}{\bar\mu\tau^2}\right]+\frac{3}{2}\xi\theta+\frac{6}{\tau^{2/3}}.
\end{equation}
\noindent From Eqs. (20), (27) and (28), we find the following form of metric functions in terms of $\tau$
\begin{equation}
A=\tau^{\frac{1}{3}},
\end{equation}
\begin{equation}
B=\frac{1}{\sqrt{d_1}}\;\tau^{\frac{1}{3}}\exp{\left[-\frac{k_1}{2}\int\frac{dt}{\tau}\right]},
\end{equation}
\begin{equation}
C=\sqrt{d_1}\;\tau^{\frac{1}{3}}\exp{\left[\frac{k_1}{2}\int\frac{dt}{\tau}\right]}.
\end{equation}
\noindent Now, Eqs. (24) and (29) take the forms
\begin{equation}
  \dot\rho+\left(\rho-\frac{\lambda}{3}\right)\frac{\dot\tau}{\tau}=\xi\frac{\dot\tau^2}{\tau^2},
\end{equation}
\noindent and
\begin{equation}
\ddot\tau=\frac{1}{2}(3\rho+\lambda)\tau+\frac{k}{2}\;\tau^{-\frac{1}{3}}+6\tau^\frac{1}{3}+\frac{3}{2}\xi\theta\tau,
\end{equation}
\noindent where $k=\frac{I^2}{\bar\mu}$.\\
\indent Consider the equation of state for a cloud of string models [6]
 \begin{equation}
\rho=\alpha\lambda,
\end{equation}
\noindent where the constant $\alpha$ is defined as
 \begin{eqnarray}
 \nonumber
\alpha&=&1 \;\;\;\;\;(\text{geometric or\; Nambu string}),\nonumber\\
&=&(1+\omega)\;\;\; (\text{p- string or Takabayasi string}),\nonumber\\
&=&-1 \;\;\;(\text{Reddy string}),
\end{eqnarray}
\noindent where $\omega$ is a positive constant. We further assume that the coefficient of bulk viscosity is inversely proportional to the expansion scalar [70], i.e.,
\begin{equation}
  \xi\theta=k_2,
\end{equation}
\noindent where $k_2$ is a positive constant. From (33) and (35), we get
 \begin{equation}
\frac{\dot\rho}{\left(1-\frac{1}{3\alpha}\right)\rho-k_2}=-\frac{\dot\tau}{\tau},
\end{equation}
\noindent which on integration, gives
\begin{equation}
\rho=\frac{3\alpha}{\left(3\alpha-1\right)}\left[k_2+k_3\tau^{-\left(\frac{3\alpha-1}{3\alpha}\right)}\right],
\end{equation}
\noindent where $k_3$ is a constant of integration. For $\rho>0$, we must have either $\alpha>1/3$ or $\alpha <0$. It means that all the above three string models (36), may be described by assuming the relation (37). Further, inserting $\rho$ from (39) into (34), we find \\
\begin{equation}
\ddot\tau=\frac{3\left(3\alpha+1\right)k_3}{2\left(3\alpha-1\right)}\tau^{\frac{1}{3\alpha}}+\frac{k}{2}\tau^{-\frac{1}{3}}
+6\tau^{\frac{1}{3}}+\frac{1}{2}\left(\frac{18\alpha}{3\alpha-1}\right)k_2\tau,
\end{equation}
\noindent whose solution is
\begin{equation}
\dot\tau^2=\left(\frac{9\alpha k_3}{3\alpha-1}\right)\tau^{\frac{3\alpha+1}{3\alpha}}
+\frac{3}{2}k\tau^{\frac{2}{3}}+9\tau^{\frac{4}{3}}+\left(\frac{9\alpha k_2}{3\alpha-1}\right)\tau^2+k_4,
\end{equation}
\noindent where $k_4$ is a constant of integration. Now, Eq. (41) can be rewritten as
\begin{equation}
\dot\tau=\sqrt{\left(\frac{9\alpha k_3}{3\alpha-1}\right)\tau^{\frac{3\alpha+1}{3\alpha}}
+\frac{3}{2}k\tau^{\frac{2}{3}}+9\tau^{\frac{4}{3}}+\left(\frac{9\alpha k_2}{3\alpha-1}\right)\tau^2+k_4}.
\end{equation}
\indent Taking into account that the energy density and string tension density obey the equation of state (35), we conclude that $\rho$ and $\lambda$, i.e., the right hand side of Eq.(34) is a function of $\tau$ only, i.e.,
\begin{equation}
\ddot\tau=F(\tau).
\end{equation}
From the mechanical point of view, Eq. (43) can be interpreted as equation of motion of a single particle with unit mass under the force $F(\tau)$. Then, the following first integral exists
\begin{equation}
\dot\tau=\sqrt{2[\epsilon-u(\tau)]},
\end{equation}
\noindent where $\epsilon$ can be viewed as energy level and $u(\tau)$ is the potential of the force $F$. A comprehensive description concerning potential is found in ref. [71]. Comparing (42) and (44), we find $\epsilon=k_4/2$ and
\begin{equation}
u(\tau)=-\frac{1}{2}\left[\left(\frac{9\alpha k_3}{3\alpha-1}\right)\tau^{\frac{3\alpha+1}{3\alpha}}
+\frac{3}{2}k\tau^{\frac{2}{3}}+9\tau^{\frac{4}{3}}+\left(\frac{9\alpha k_2}{3\alpha-1}\right)\tau^2\right].
\end{equation}
\noindent Finally, we write Eq. (42) in a general quadrature form as
\begin{equation}
\int\frac{d\tau}{\sqrt{\left(\frac{9\alpha k_3}{3\alpha-1}\right)\tau^{\frac{3\alpha+1}{3\alpha}}
+\frac{3}{2}k\tau^{\frac{2}{3}}+9\tau^{\frac{4}{3}}+\left(\frac{9\alpha k_2}{3\alpha-1}\right)\tau^2+k_4}}=t+t_0,
\end{equation}
\noindent where the integration constant $t_0$ can be taken as zero for simplicity.\\

\begin{center}
\noindent \textbf{4. Solution of various string models}\\
\end{center}

\indent We observe that it is too difficult to solve (46), in general. Therefore, we present the following three string models depending on the values of $\alpha$ as defined in (36).\\

\noindent \textbf{4.1 Geometric string model ($\alpha=1$) }\\

Let us find the solution for viscous fluid with and without magnetic field in the following subsections.\\

\noindent \textbf{4.1.1 Viscous fluid solution with magnetic field}\\

\noindent For $k_4=0$, Eq. (46) reduces to
\begin{equation}
\int\frac{d\tau}{\sqrt{\frac{9}{2}\left(k_3+2\right)\tau^{\frac{4}{3}}+\frac{3}{2}k\tau^\frac{2}{3}+\frac{9}{2}k_2\tau^2}}=t.
\end{equation}
\noindent For $3(k_3+2)^2<4kk_2$, on integration Eq.(47),  gives
\begin{equation}
  \tau=\left[\frac{1}{2k_2}\sqrt{\frac{4kk_2-3(k_3+2)^2}{3}}\;\sinh{\left(\sqrt{2k_2}\;t\right)}-
  \frac{k_3+2}{2k_2}\right]^\frac{3}{2}, \;\;\;\;\; k_2\neq0,
\end{equation}
\noindent and when $3(k_3+2)^2>4kk_2$, Eq.(47) gives
\begin{equation}
  \tau=\left[\frac{1}{2k_2}\sqrt{\frac{3(k_3+2)^2-4kk_2}{3}}\;\cosh{\left(\sqrt{2k_2}\;t\right)}-
  \frac{k_3+2}{2k_2}\right]^\frac{3}{2},\;\;\;\;\;k_2\neq0,
\end{equation}

\noindent For small $t$, we have $\sinh{\left(\sqrt{2k_2}\;t\right)}\approx\sqrt{2k_2}\;t$. Therefore, Eq. (48) can be written as
\begin{equation}
  \tau=(P_1t-Q_1)^{3/2},
\end{equation}
\noindent where $P_1=\sqrt{\frac{4kk_2-3(k_3+2)^2}{6k_2}}$ and $Q_1=\frac{k_3+2}{2k_2}$. \\
\indent At $t=0$, $\tau$ becomes imaginary. Thus, for reality of the model, $t$ must satisfy $t>\frac{Q_1}{P_1}$.\\
\indent Using (50), Eqs.(30)-(32) take the forms
\begin{equation}
  A=\sqrt{P_1t-Q_1},
\end{equation}
\begin{equation}
  B=\frac{1}{\sqrt{d_1}}\sqrt{P_1t-Q_1}\;\exp{\left[\frac{k_1}{P_1\sqrt{P_1t-Q_1}}\right]},
\end{equation}
\begin{equation}
  C=\sqrt{d_1}\sqrt{P_1t-Q_1}\;\exp{\left[-\frac{k_1}{P_1\sqrt{P_1t-Q_1}}\right]}.
\end{equation}
\noindent The directional Hubble parameters along $x$, $y$ and $z$ axes are respectively given by
\begin{equation}
  H_1=\frac{P_1}{2(P_1t-Q_1)},\;\;H_2=\frac{P_1}{2(P_1t-Q_1)}-\frac{k_1}{2(P_1t-Q_1)^\frac{3}{2}},\;\; H_3=\frac{P_1}{2(P_1t-Q_1)}+\frac{k_1}{2(P_1t-Q_1)^\frac{3}{2}}.
\end{equation}
\noindent The average Hubble parameter in terms of cosmic time $t$ becomes
\begin{equation}
  H=\frac{P_1}{2(P_1t-Q_1)}.
\end{equation}
\noindent The anisotropic parameter and shear scalar, respectively have the following expressions
\begin{equation}
\Delta=\frac{2k_1^2}{3P_1^2(P_1t-Q_1)},\;\;\sigma^2=\frac{k_1^2}{4(P_1t-Q_1)^3}.
\end{equation}
\noindent We observe that the scale factors increase with time for $t>\frac{Q_1}{P_1}$. The other physical parameters (54)- (56) diverse at $t=\frac{Q_1}{P_1}$ and tend to zero as $t \rightarrow\infty$. From (55) and (56), we get
\begin{equation}
\frac{\sigma}{\theta}=\frac{k_1}{3P_1\sqrt{(P_1t-Q_1)}},
\end{equation}
\noindent which is time-dependent and tends to zero as $t\rightarrow\infty$, which shows that the model becomes isotropic in late time. The deceleration parameter becomes $q=1$, i.e.,  a positive constant. Therefore, the present model expands with the decelerated rate throughout the evolution.\\
\indent The energy density and string tension density are given by
\begin{equation}
  \rho=\lambda= \frac{3}{2}\left(k_2+\frac{k_3}{P_1t-Q_1}\right),
\end{equation}
\noindent which shows that the matter behaves as a cloud of geometric strings. The particle density, $\rho_p$ remains zero throughout the evolution. We observe that $\rho$ and $\lambda$  remains positive throughout the evolution and become infinite at the initial epoch at $t=\frac{Q_1}{P_1}$. However, $\rho$ and $\lambda$ decrease with time for $t>\frac{Q_1}{P_1}$ and approach to a constant value, $\frac{3}{2}k_2$ as $t\rightarrow\infty$. We observe that this constant value of $\rho$ is due to the viscous term, $k_2$. It means that the viscosity parameter prevents the universe to be empty at late times of its evolution. \\
\indent The classical potential,(45) in terms of $t$ takes the form
\begin{equation}
  u(t)=-\frac{3}{4}\left[3k_2(P_1t-Q_1)^3+3(k_3+2)(P_1t-Q_1)^2+k(P_1t-Q_1)\right].
\end{equation}
\noindent The behavior of the classical potential has been shown in figure 1.\\

\noindent \textbf{4.1.2 Viscous fluid solution without magnetic field}\\

\noindent  For $\alpha=1$ and in the absence of magnetic field ($k=0$), Eq.(46) becomes
\begin{equation}
\int\frac{d\tau}{\sqrt{\frac{9}{2}\left(k_3+2\right)\tau^{\frac{4}{3}}+\frac{9}{2}k_2\tau^2}}=t,
\end{equation}
\noindent which on integration it gives
\begin{equation}
  \tau=\left[\frac{(k_3+2)}{2k_2}\left(\cosh(\sqrt{2k_2}\;t)-1\right)\right]^\frac{3}{2},\;\;\;\;\;k_2\neq0,
\end{equation}
\noindent We find that the solution (49) also gives the same solution (61) in the absence of magnetic field, i.e., $k=0$. Therefore, the solution (49) may be considered for viscous fluid without magnetic field.\\
\indent For small $t$, Eq.(61) gives
\begin{equation}
\tau=\left(\frac{k_3+2}{2}\right)^\frac{3}{2}\;t^3.
\end{equation}
Therefore, the scale factors in terms of $t$ have the following solutions
\begin{equation}
A=\sqrt{\left(\frac{k_3+2}{2}\right)}\;t,
\end{equation}
\begin{equation}
B=\frac{1}{\sqrt{d_1}}\sqrt{\left(\frac{k_3+2}{2}\right)}\;t \exp \left[\frac{k_1}{4}\left(\frac{2}{k_3+2}\right)^{\frac{3}{2}}\;{\frac{1}{t^{2}}}\right],
\end{equation}
\begin{equation}
C=\sqrt{d_1}\sqrt{\left(\frac{k_3+2}{2}\right)}\;t \exp \left[-\frac{k_1}{4}\left(\frac{2}{k_3+2}\right)^{\frac{3}{2}}\;{\frac{1}{t^{2}}}\right].
\end{equation}
\noindent The directional Hubble parameters along $x$, $y$ and $z$ axes are respectively given by
\begin{equation}
  H_1=\frac{1}{t},\;\;\;\;\ H_2=\frac{1}{t}-\frac{k_1}{4}\left(\frac{2}{k_3+2}\right)^{\frac{3}{2}}\;\frac{1}{t},\;\;\;
  H_3=\frac{1}{t}+\frac{k_1}{4}\left(\frac{2}{k_3+2}\right)^{\frac{3}{2}}\;\frac{1}{t}.
\end{equation}
The average Hubble parameter is
\begin{equation}
H=\frac{1}{t}
\end{equation}
The anisotropy parameter and shear scalar are respectively take the form
\begin{equation}
\Delta=\frac{k_1^2}{3(k_3+2)^3},\;\;\;\; \sigma^{2}=\frac{k_1^2}{9(k_3+2)^3}\;{\frac{1}{t^2}}.
\end{equation}
\noindent From above solutions we observe that the universe starts with $t=0$ and expands in all directions for all $t>0$. The rate of expansion in each direction diverse at $t=0$ and tends to zero as $t\rightarrow\infty$. The anisotropy parameter becomes constant throughout the evolution of the universe where as the shear scalar varies inversely with the square of time. The model remains anisotropic throughout the evolution as the ratio of $\sigma/\theta=1/9(k_3+2)^{3/2}$, i.e., a constant. The deceleration parameter $q$ is zero which shows margin inflation during the expansion.\\
\indent The energy density and string tension density are given by
\begin{equation}
  \rho=\lambda= \frac{3}{2}\left(k_2+\frac{2k_3}{k_3+2}\;\frac{1}{t^2}\right),
\end{equation}

\begin{figure}[h]
\begin{center}
\includegraphics[width=8 cm, height=5 cm]{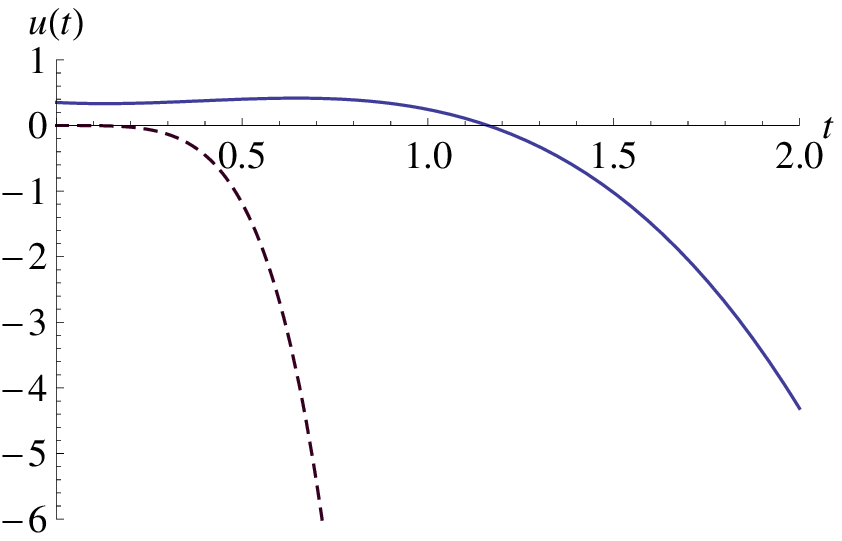}\\
{\footnotesize \textbf{Fig. 1}\;\; Classical Potential versus time for viscous fluid solution with (bold line) and without (dashed line) magnetic field.}
 \end{center}
\end{figure}
We find that the energy density remains positive through out the evolution of the universe. It is infinite at initial epoch and decreases with time, and ultimately attains a constant value, $\frac{3k_2}{2}$ in late time expansion of the universe. Therefore, the bulk viscosity prevents the universe to be empty during late time of evolution. From (58) and (69) we find that the energy density attains the same constant value with and without magnetic field in late time of evolution.\\
\indent The classical potential (45) in terms of $t$ takes the form
\begin{equation}
  u(t)=-\frac{9(k_3+2)^3}{32}(2+k_2t^2)t^4.
\end{equation}
Figure 1 plots the potential with respect to time in the presence of viscous fluid and magnetic field (bold line) and with viscous fluid only (dashed line). We have used the numerical value of various constants as $k_2=2$, $k_3=1$, $I=2$ or $0$ and $\bar\mu=1.00001$. We observe that $\mu(t)$ shows positive and negative nature with respect to time $t$ in magnetized viscous fluid. However, it has always negative value and decreases rapidly with time in the presence of viscous fluid only. \\

\noindent \textbf{4.2. Reddy string model ($\alpha=-1$) }\\

Let us find the solution for viscous fluid with and without magnetic field in the following subsections for Reddy string model.\\

\noindent \textbf{4.2.1 Viscous fluid solution with magnetic field}\\

\noindent For $k_4=0$, Eq.(46) becomes
\begin{equation}
\int\frac{d\tau}{\sqrt{9\tau^\frac{4}{3}+\frac{3}{4}(3k_3+2k)\tau^{\frac{2}{3}}+\frac{9}{4}k_2\tau^2}}=t.
\end{equation}
\noindent For $k_2\left(3k_3+2k\right)<12$, on integration (71), we obtain
\begin{equation}
  \tau=\left[\frac{1}{k_2}\sqrt{\frac{12-k_2(3k_3+2k)}{{3}}}\;\cosh{\left(\sqrt{k_2}\;t\right)}-
  \frac{2}{k_2}\right]^\frac{3}{2},
\end{equation}
\noindent and for $k_2\left(3k_3+2k\right)>12$, Eq. (71) gives
\begin{equation}
  \tau=\left[\frac{1}{k_2}\sqrt{\frac{k_2(3k_3+2k)-12}{{3}}}\;\sinh{\left(\sqrt{k_2}\;t\right)}-
  \frac{2}{k_2}\right]^\frac{3}{2},
\end{equation}
\noindent For small $t$, we have $\sinh{\left(\sqrt{k_2}\;t\right)}\approx\sqrt{k_2}\;t$, and therefore, (73) takes the form
\begin{equation}
  \tau=(P_2t-Q_2)^{3/2},
\end{equation}
\noindent where $P_2=\sqrt{\frac{1}{3k_2}[k_2(3k_3+2k)-12]}$ and $Q_2=\frac{2}{k_2}$.\\
\indent At $t=0$, $\tau$ becomes imaginary. For reality of the model, $t$ must satisfy $t>\frac{Q_2}{P_2}$. From Eqs.(30)-(32) and (74), we find
\begin{equation}
  A=\sqrt{P_2t-Q_2},
\end{equation}
\begin{equation}
  B=\frac{1}{\sqrt{d_1}}\sqrt{P_2t-Q_2}\;\exp{\left[\frac{k_1}{P_2\sqrt{P_2t-Q_2}}\right]},
\end{equation}
\begin{equation}
  C=\sqrt{d_1}\sqrt{P_2t-Q_2}\;\exp{\left[-\frac{k_1}{P_2\sqrt{P_2t-Q_2}}\right]}.
\end{equation}
\noindent The directional Hubble parameters along $x$, $y$ and $z$ axes are respectively given by
\begin{equation}
  H_1=\frac{P_2}{2(P_2t-Q_2)},\;\;H_2=\frac{P_2}{2(P_2t-Q_2)}-\frac{k_1}{2(P_2t-Q_2)^\frac{3}{2}},\;\; H_3=\frac{P_2}{2(P_2t-Q_)}+\frac{k_1}{2(P_2t-Q_2)^\frac{3}{2}}.
\end{equation}
\noindent The average Hubble parameter in terms of cosmic time $t$ is
\begin{equation}
  H=\frac{P_2}{2(P_2t-Q_2)}.
\end{equation}
\noindent The anisotropic parameter and shear scalar, respectively take the forms
\begin{equation}
\Delta=\frac{2k_1^2}{3P_2^2(P_2t-Q_2)},\;\;\sigma^2=\frac{k_1^2}{4(P_2t-Q_2)^3}.
\end{equation}
\noindent The above physical parameters in Eqs. (78)-(80) diverge at $t=\frac{Q_2}{P_2}$ and asymptotically tend to zero as $t\to\infty$. From (79) and (80), we get
\begin{equation}
\frac{\sigma}{\theta}=\frac{k_1}{3P_2\sqrt{(P_2t-Q_2)}}.
\end{equation}
\noindent which is time dependent and tends to zero as $t\to\infty$. Therefore, this string model also becomes isotropic for large $t$. The energy density, string tension density and particle density are respectively given by
\begin{equation}
  \rho= \frac{3}{4}\left[k_2+\frac{k_3}{(P_2t-Q_2)^2}\right],\;\; \lambda= -\frac{3}{4}\left[k_2+\frac{k_3}{(P_2t-Q_2)^2}\right],\;\;  \rho_p= \frac{3}{2}\left[k_2+\frac{k_3}{(P_2t-Q_2)^2}\right].
\end{equation}
 \noindent We find that $\rho$ and $\rho_p$ are infinite at $t=\frac{Q_2}{P_2}$ and tend to a constant values, $\frac{3k_2}{4}$ and $\frac{3k_2}{2}$, respectively as $t\to \infty$. The constant values of $\rho$ and $\rho_p$ in the later stages of evolution are due to the bulk viscosity. \\
\indent The string tension density $\lambda$ remains negative and gradually increases with time and finally it approaches to a constant value as $t\to\infty$. \\
\indent The potential in terms of $t$ is written as
\begin{equation}
  u(t)=-\frac{3}{8}\left[3k_2(P_2t-Q_2)^3+12(P_2t-Q_2)^2+(3k_3+2k)(P_2t-Q_2)\right].
\end{equation}
The nature of different density parameters and $u(t)$ have been shown in figures 2 and 3, respectively.\\

\noindent \textbf{4.2.2 Viscous fluid solution without magnetic field}\\

\noindent For $\alpha=-1$ and in the absence of magnetic field ($k=0$), Eq.(46) gives
\begin{equation}
\int\frac{d\tau}{\sqrt{9\tau^\frac{4}{3}+\frac{9k_3}{4}\tau^{\frac{2}{3}}+\frac{9}{4}k_2\tau^2}}=t.
\end{equation}
\noindent For $k_2k_3<4$, on integration (84), we obtain
\begin{equation}
  \tau=\left[\frac{1}{k_2}\sqrt{4-k_2k_3}\;\cosh{\left(\sqrt{k_2}\;t\right)}-
  \frac{2}{k_2}\right]^\frac{3}{2},
\end{equation}
\noindent and for $k_2k_3>4$, Eq.(84) gives
\begin{equation}
  \tau=\left[\frac{1}{k_2}\sqrt{k_2k_3-4}\;\sinh{\left(\sqrt{k_2}\;t\right)}-
  \frac{2}{k_2}\right]^\frac{3}{2},
\end{equation}
One may observe that the above solutions (85) and (86) are same as directly obtained by putting $k=0$ in (72) and (73), respectively. Therefore, the solution of various physical parameters and their physical significance of this viscous solution  may be discussed in the line of subsection 4.2.1 by taking $k=0$.\\
\indent For small $t$ , Eq.(86) gives
\begin{equation}
\tau=\left[\sqrt{\frac{k_2k_3-4}{k_2}}\;t-\frac{2}{k_2}\right]^{\frac{3}{2}},
\end{equation}
which is same as the solution (74) in the absence of magnetic field ($k=0$). \\
\begin{figure}[h]
\begin{center}
\includegraphics[width=8 cm, height=5 cm]{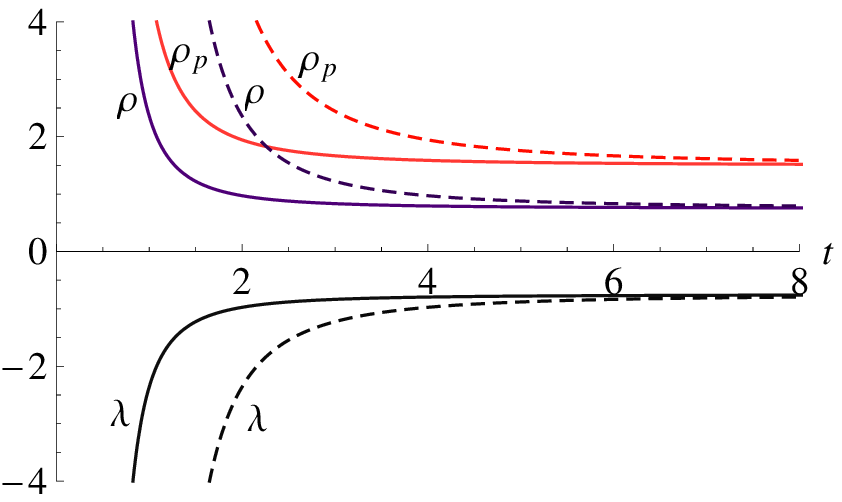}\\
 {\footnotesize \textbf{Fig. 2}\;\;Density parameters versus time  for viscous fluid solution with (bold lines) and without (dashed lines) magnetic field.}
 \end{center}
\end{figure}

\begin{figure}[h]
\begin{center}
\includegraphics[width=8 cm, height=5 cm]{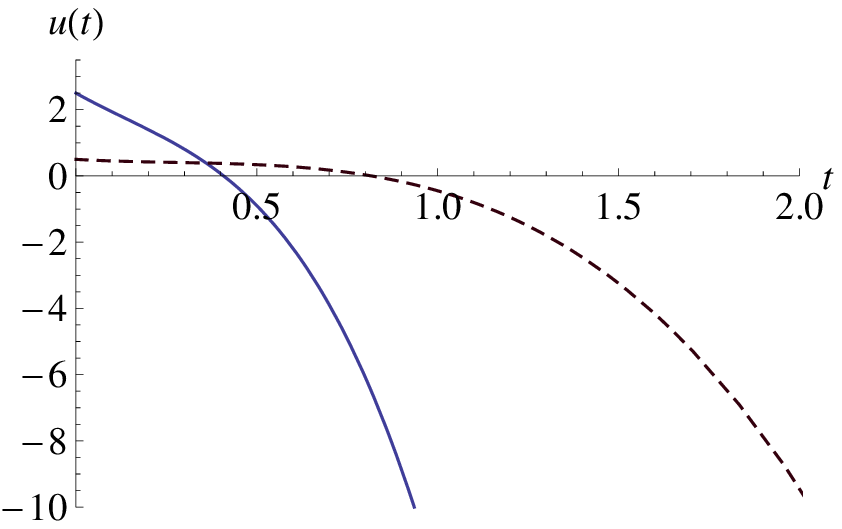}\\
 {\footnotesize \textbf{Fig. 3}\;\; Potential versus time for viscous fluid solution with (bold lines) and without (dashed lines) magnetic field.}
 \end{center}
\end{figure}
\indent Figures 2 plot the graph of energy density, energy tension density and particle density with respect to time in presence of magnetized viscous fluid (bold lines) and in presence of viscous only (dashed lines) for $k_2=1$, $k_3=2$, $I=2$ and $\bar\mu=1.00001$. The figure shows that the energy density and particle density are positive through out the evolution and decrease with time where as the energy tension density is negative and increases with time and attains a constant value in late time in both cases. It is interesting to note that these parameters attain the same respective value in both the cases in late time evolution.
\noindent The behavior of classical potential with time in both cases is shown in fig.3 for $k_2=3$, $k_3=2$, $I=2$ or $0$ and $\bar\mu=1.00001$. The figure shows that $\mu(t)$ is positive and negative in nature in both cases but it decreases rapidly in former case. \\

\noindent \textbf{4.3 Takabayashi string model $(\alpha=1+\omega)$}\\

\noindent For $k_4=0$, Eq.(46) reduces to
\begin{equation}
\int\frac{d\tau}{\sqrt{\frac{9(1+\omega)k_3}{3\omega+2}\tau^{\frac{3\omega+4}{3\omega+3}}
+\frac{3k}{2}\tau^{\frac{2}{3}}+9\tau^{\frac{4}{3}}+\frac{9(1+\omega)k_2}{3\omega+2}\tau^2}}=t.
\end{equation}
\noindent One can observe that it is very difficult to find a general solution of $\tau$ in terms of $t$. Therefore, we express $\rho$, $\lambda$ and $\rho_p$ in terms of $\tau$ as
\begin{equation}
\rho=\frac{3(1+\omega)}{3\omega+2}\left[k_2+k_3\tau^{-\frac{2+3\omega}{3(1+\omega)}}\right],
\end{equation}
\begin{equation}
\lambda=\frac{3}{3\omega+2}\left[k_2+k_3\tau^{-\frac{2+3\omega}{3(1+\omega)}}\right],
\end{equation}
\begin{equation}
\rho_{p}=\frac{3\omega}{3\omega+2}\left[k_2+k_3\tau^{-\frac{2+3\omega}{3(1+\omega)}}\right],
\end{equation}
\noindent As $\omega>0$, we find that $\rho$, $\lambda$ and $\rho_p$ remain positive through out the evolution of the universe. These physical parameters are decreasing function of $\tau$ and become constant for large $t$ due to the bulk viscosity. The geometrical string model may be recovered for $\omega=0$ as discussed in Sect. 4.1. The classical potential in this is given by
\begin{equation}
u(\tau)=-\frac{3}{2}\left[\frac{3k_3(1+\omega)}{3\omega+2}\tau^{\frac{4+3\omega}{3(1+\omega)}}
+\frac{k}{2}\tau^{\frac{2}{3}}+3\tau^{\frac{4}{3}}+\frac{3k_2(1+\omega)}{3\omega+2}\tau^2\right].
\end{equation}
\begin{figure}[h]
\begin{center}
\includegraphics[width=8 cm, height=5 cm]{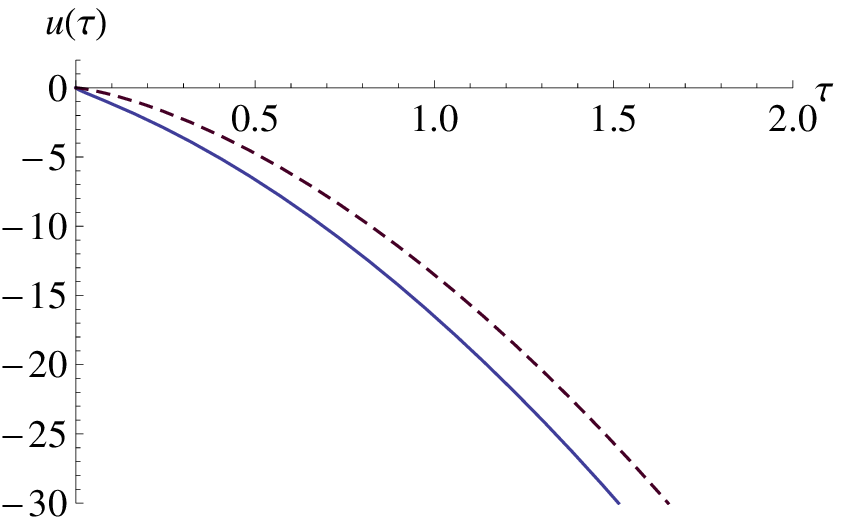}\\
{\footnotesize \textbf{Fig. 4}\;\;  Potential versus time for viscous fluid solution with (bold lines) and without (dashed lines) magnetic field.}
 \end{center}
\end{figure}
\noindent Fig. 4 illustrates the behavior of potential $\mu(t)$ with respect to $\tau$ for $k_2=3$, $k_3=2$, $I=2$, or $0$, $\bar\mu=1.00001$ and $\omega=1$. It is clear that potential remains negative and decreases rapidly throughout the evolution of the universe.\\

\begin{center}
\noindent \textbf{5. Conclusion}\\
\end{center}

\indent We have studied anisotropic Bianchi V string cosmological model with viscous fluid and magnetic field in general relativity by taking certain physical assumptions. Since viscous fluid and magnetic field have cosmological origin, it is interesting to discuss the viscous and magnetic field effects on the expansion history of the universe in early and late stages of evolution in string cosmology. The Einstein's field equations have been solved exactly for geometrical and Reddy string models for viscous fluid with and without magnetic field whereas a general quadrature form of average scale factor has been found in Takabayashi string model. The solutions present interesting features in the presence of viscous fluid and magnetic field and in the presence of viscous fluid only. The effect of viscous term has more important for the expansion of the universe as we have already discussed in different string models. We have also plotted the graphs of various physical parameters to show the effects of magnetized viscous fluid and bulk viscous fluid without magnetic field.\\
 \indent In geometrical string model we have observed that proper energy density remains positive through out the evolution and attains the same constant value during late time for viscous fluid with and without magnetic field. Hence, the presence of viscous term prevents to be empty in its future evolution. The classical potential changes its behavior rapidly due to the bulk viscous term. It is negative through out the evolution in the absence of magnetic field but it is positive for some finite time and after that it shows the negative nature during late time. \\
 \indent In Reddy string model the proper energy density and particle density remain positive through out the evolution and attains the same constant value during late time for viscous fluid with and without magnetic field. The string tension density is  always negative and increases with time, and turns out to be finite during late time.  At initial epoch, these physical parameters are infinite while they attain to be finite in late time due to bulk viscous effects in both cases. \\
 \indent The other physical and kinematical parameters such as the respective directional Hubble parameters,  anisotropic parameter and shear scalar $\sigma$ in both the string models are infinite at initial epoch and approach to zero asymptotically. The models approach isotropic at later stages of the evolution of the universe for viscous fluid with magnetic field where as these models are always anisotropic for viscous fluid without magnetic field. The deceleration parameter turns out to be a positive constant in each string model in the presence of viscous fluid with magnetic field which shows that the universe expands with decelerated rate. It is zero in case of viscous fluid only showing the marginal inflation during the expansion. \\
 \indent In Takabayashi model we have found a quadrature form of average volume which is too difficult to solve, in general. The solution for the  proper energy density, string tension density and particle energy density in terms of average scale factor have been represented. These physical parameters approach to a constant value asymptotically as $t\rightarrow\infty$ due to the bulk viscosity.\\
\indent We have also discussed the classical potential with respect to time in each string model and have observed that the classical potential changes its behavior rapidly due to the bulk viscous term. In geometrical string model it remains negative and decreases rapidly in viscous fluid solution without magnetic field. In Takabayashi string model it remains negative in both cases but the effect of viscous fluid with and without magnetic field vary rapidly. Thus, we conclude that the bulk viscous fluid with and without magnetic field plays an important role in the evolution of the universe in anisotropic models. Some more interesting properties of these string models for viscous fluid with and without solution will be discussed elsewhere.\\

\begin{center}
\textbf{References}
\end{center}
\noindent [1] T.W.B. Kibble, J. Phys. A {\bf9}, 1387 (1976).\\
\noindent [2] A.E. Everett, Phys. Rev. D {\bf24}, 858 (1981).\\
\noindent [3] A. Vilenkin,  Phys. Rev. D {\bf24}, 2082 (1981).\\
\noindent [4] P.S. Letelier, Phys. Rev. D {\bf20}, 1294 (1979).\\
\noindent [5] J. Stachel, Phys. Rev. D {\bf21}, 2171 (1980).\\
\noindent [6] P.S. Letelier,  Phys. Rev. D {\bf28}, 2414 (1983).\\
\noindent [7] D.R. Matraverse,  Gen. Relativ. Grav. {\bf20}, 279 (1988).\\
\noindent [8] K.D. Krori, T. Chaudhury, C.R. Mahanta, A. Mazumdar, Gen. Relativ. Grav. {\bf22}, 123 \indent(1990).\\
\noindent [9] I. Yavuz and I. Tarhan,  Astrophys. Space Sci. {\bf240}, 45 (1996).\\
\noindent [10] R. Bali and S. Dave, Pramana J. Phys. {\bf56}, 513 (2001).\\
\noindent [11] R. Bali and S. Dave,  Astrophys. Space Sci. {\bf288}, 503 (2003).\\
\noindent [12] R. Bali and R.D. Upadhaya, Astrophys. Space Sci. {\bf283}, 97 (2003).\\
\noindent [13] R. Bali and D.K. Singh,  Astrophys. Space Sci. {\bf300}, 387 (2005).\\
\noindent [14] R. Bali and A. Pradhan, Chin. Phys. Lett. {\bf24}, 585 (2007).\\
\noindent [15] A. Pradhan and D.S. Chouhan,  Astrophys. Space Sci. {\bf331}, 697 (2011).\\
\noindent [16] K.L. Mahanto, S.K. Biswal, S.K. Sahoo and M.C. Adhikary, Int. J. Theor. Phys. {\bf51}, \\
\indent 1538 (2012).\\
\noindent [17] E. Asseo and H. Sol, Phys. Rep. {\bf6}, 148 (1987).\\
\noindent [18] M.S. Madsen,  MNRAS {\bf237}, 109 (1989).\\
\noindent [19] A.M. Wolfe, K. Lanzetta and A.L. Oren,  Astrophys. J. {\bf388}, 17 (1992).\\
\noindent [20] R. Kulsrud, R. Cen, J.P. Ostriker and D. Ryu, Astrophys. J. {\bf380}, 481 (1997).\\
\noindent [21] J.D. Barrow, Phys. Rev. D {\bf55},7451 (1997).\\
\noindent [22] D.R. Matravers and C.G. Tsagas,  Phys. Rev. D {\bf62}, 103519 (2000).\\
\noindent [23] A. Banerjee, A.K. Sanyal and S. Chakraborty,  Pramana-J. Phys. {\bf34}, 1 (1990).\\
\noindent [24] S. Chakraborty, Ind. J. Pure Appl. Phys. {\bf29}, 31 (1991).\\
\noindent [25] R. Tikekar and L.K. Patel, Gen. Relativ. Grav. {\bf24}, 397, (1992).\\
\noindent [26] R. Tikekar and L.K. Patel, Pramana J. Phys. {\bf42}, 483 (1994).\\
\noindent [27] ShriRam and J. K. Singh, Gen. Relativ. Grav. {\bf27}, 1207 (1995).\\
\noindent [28] L.K. Patel and S.D. Maharaj, Pramana J. Phys. {\bf47}, 1 (1996).\\
\noindent [29] G.P. Singh and T. Singh, Gen. Relativ. Grav. {\bf31}, 371 (1999).\\
\noindent [30] J.K. Singh and  ShriRam,  Astrophys. Sapce Sci. {\bf246}, 65 (1997).\\
\noindent [31] C. B. Kilin\c{c} and I. Yavuz,  Astrophys. Space Sci {\bf271}, 11 (2000).\\
\noindent [32] A. Pradhan, A. Rai and S.K. Singh,  Astrophys. Space Sci. {\bf312}, 261 (2007).\\
\noindent [33] A. Pradhan, Fizika B {\bf16}, 205 (2007).\\
\noindent [34] R. Bali and S. Jain,  Int. J. Mod. Phys. D {\bf16}, 1769 (2007).\\
\noindent [35] B. Saha and M. Visinescu, Astrophys. Space Sci. {\bf315}, 99 (2008).\\
\noindent [36] A. Pradhan, K. Jatonia and  A. Singh,  Braz. J. Phys. {\bf38}, 167 (2008).\\
\noindent [37] A. Pradhan, V. Rai and K. Jatonia,  Commun. Theor. Phys. {\bf50}, 279 (2008).\\
\noindent [38] B. Saha, V. Rikhvitsky and  M. Visinescu,  Cent. Eur. J. Phys. {\bf8}, 113 (2010).\\
\noindent [39] A. Pradhan, H. Amirhashchi and  H. Zainuddin,  Int. J. Thoer. Phys. {\bf50}, 56 (2011).\\
\noindent [40] H. Amirhashchi, H. Zainuddin and A. Pradhan,  Int. J. Thoer. Phys. {\bf50}, 2531 (2011).\\
\noindent [41] V. Rikhvitsky, B. Saha and M. Visinescu,  Astrophys. Sapce. Sci. {\bf339}, 371 (2012).\\
\noindent [42] M. Cataldo, N. Cruz and S. Lepe, Phys. Lett. B {\bf619}, 5 (2005).\\
\noindent [43] G.L. Murphy,  Phys. Rev. D {\bf8}, 4231 (1973).\\
\noindent [44] V.A. Belinskii and I.M. Khalatrikov, Soviet Phys. JETP {\bf42}, 205 (1976).\\
\noindent [45] W.H. Huang,  Phys. Lett. A {\bf129}, 429 (1988).\\
\noindent [46] W. Zimdahl,  Phys. Rev. D {\bf53}, 5483 (1996).\\
\noindent [47] W. Zimdahl,  Phys. Rev. D {\bf61}, 083511 (2000).\\
\noindent [48] L.P. Chimento, A.S. Jakubi, V. M\'{e}ndez and R. Maartens,  Class. Quantum Grav. {\bf14}, \\
\indent 3363 (1997).\\
\noindent [49] M.D. Mak and  T. Harko,  J. Math. Phys. {\bf39}, 5458 (1998).\\
\noindent [50] R. Maartens, Class. Quantum Grav. {\bf 12}, 1455 (1995).\\
\noindent [51] C.P. Singh,  Pramana J. Phys.{\bf71}, 33 (2008).\\
\noindent [52] M.K. Yadav, A. Rai and A. Pradhan,  Int. J. Theor. Phys. {\bf46}, 2677 (2007).\\
\noindent [53] G. Mohanty and C.S. Gauranga,  Tur. J. Phys {\bf32}, 251 (2008).\\
\noindent [54] S.K. Tripathy and D. Behera,  Astrophys. Space Sci. {\bf330}, 191 (2010).\\
\noindent [55] A.A. Coley,  Gen. Relativ. Grav. {\bf22}, 3 (1990).\\
\noindent [56] C.P. Singh, M. Zeyauddin and ShriRam,  Astrophys. Space Sci. {\bf315}, 181 (2008).\\
\noindent [57] C.P. Singh, M. Zeyauddin and ShriRam,  Int. J. Mod. Phys. A {\bf23}, 2719 (2008).\\
\noindent [58] C.P. Singh, M. Zeyauddin and ShriRam, Int. J. Theor. Phys. {\bf47}, 3162 (2008).\\
\noindent [59] C.P. Singh and A. Beesham,  Int. J. Mod. Phys. A {\bf25}, 3825 (2010).\\
\noindent [60] C.P. Singh,  Braz. J. Phys. {\bf41}, 323 (2011).\\
\noindent [61] N.C. Chakraborty and S. Chakraborty,  Int. J. Mod. Phys. D {\bf10}, 723 (2001).\\
\noindent [62] R. Bali,  Eur. J. Theor. Phys. {\bf5}, 105 (2008).\\
\noindent [63] R. Bali, U.K. Parrek and A. Pradhan,  Chin. Phys. Lett. {\bf24}, 2455 (2007).\\
\noindent [64] R. Bali, R. Banerjee and S.K. Banerjee,  Astrophys. Space Sci. {\bf317}, 21 (2008).\\
\noindent [65] M. Sharif and S. Waheed,  Int. J. Mod. Phys. D {\bf55}, 21500 (2012).\\
\noindent [66] C.P.Singh, Astrophys. Space Sci., DOI:10.1007/s10509-012-1236-x, (2012).\\
\noindent [67] L.D. Landuam and E.M. Lifshitz, Fluid Mechanics, Pergamon Press, Oxford, p.505 (1936).\\
\noindent [68] A. Linchnerowicz,   Relativistic Hydrodynamics and Magneto Hydrodynam\\
\indent ics, Benjamin, New York, p.13 (1967).\\
\noindent [69] R. Maartens,  Pramana J. phys. {\bf55}, 575 (2000).\\
\noindent [70] B. Saha,  Mod. Phys. Lett. A {\bf20}, 2127 (2005).\\
\noindent [71] B. Saha and T. Boyadjiev,  Phys. Rev. D {\bf69}, 124010 (2004).\\
\end{document}